\documentclass[showpacs,preprintnumbers,superscriptaddress]{revtex4}
\usepackage{CJK}
\usepackage{amsmath,amssymb,graphicx,bm}
\begin{document}

\title{Gravitational waves in conformal gravity}

\author{Rongjia Yang \footnote{Corresponding author}}
\email{yangrongjia@tsinghua.org.cn}
\affiliation{College of Physical Science and Technology, Hebei University, Baoding 071002, China}
\affiliation{Hebei Key Lab of Optic-Electronic Information and Materials, Hebei University, Baoding 071002, China}

\begin{abstract}
We consider the gravitational radiation in conformal gravity theory. We perturb the metric from flat Mikowski space and obtain the wave equation after introducing the appropriate transformation for perturbation. We derive the effective energy-momentum tensor for the gravitational radiation, which can be used to determine the energy carried by gravitational waves.
\end{abstract}


\maketitle

\section{Introduction}
The detection of gravitational waves (GWs) by the LIGO Collaboration is a milestone in GW research and opens a new window to probe general relativity (GR) and astrophysics \cite{Abbott:2016blz, Abbott:2017oio,GBM:2017lvd, Li:2017iup}. Future space-borne detectors will offer access to an unprecedented signal sensitivity \cite{AmaroSeoane:2012km}, thus it is worthwhile to explore GWs in alternative theories of gravity. Gravitational wave were considered in $f(R)$ theories \cite{Yang:2011cp, Liang:2017ahj, Rizwana:2016qdq, Capozziello:2008rq, Corda:2007nr, Corda:2007hi, Corda:2010zza, Alves:2009eg, Alves:2010ms, Berry:2011pb, Alves:2016iks, Kausar:2017uib}, in scalar-tensor theories \cite{Capozziello:2006ra, Hou:2017bqj, Zhang:2017srh}, in $f(T)$ theories \cite{Bamba:2013ooa} and in fourth-order gravity \cite{Capozziello:2015nga}. The evolution equation for gravitational perturbation in four dimensional spacetime in presence of a spatial extra dimension has been derived in \cite{Chakraborty:2017qve}. The linear perturbation of higher-order gravities has been discussed in \cite{Bueno:2016ypa}. Following the original work by Weyl \cite{Weyl1918Reine} (for review, see \cite{Scholz:2011za}), conformal gravity (CG), as a possible candidate alternative to GR, attracts much attention. It can give rise to an accelerated expansion \cite{Mannheim:2005bfa}. It was tested with astrophysical observations and had been confirmed that it does not suffer from an age problem \cite{Yang:2013skk}. It can describe the rotation curves of galaxies without dark matter \cite{OBrien:2011vks}. Cosmological perturbations in CG were investigated in \cite{Mannheim:2011is, Amarasinghe:2018jkv}. The particle content of linearized conformal gravity was considered in \cite{RIEGERT1984110}. It had been shown that CG accommodates well with currently available SNIa and GRB samples \cite{Diaferio:2011kc, Varieschi:2014pca, Roberts:2017nkm}. A series of dynamical solutions in CG were found in \cite{Zhang:2017amt}. Mass decomposition of the lens galaxies of the Sloan Lens Advanced Camera for Surveys in CG was discussed in \cite{Potapov:2016pgr}. Recently it was indicated that conformal gravity can potentially test well against all astrophysical observations to date \cite{Cattani:2013dla}. It has been shown that CG can also give rise to an inflationary phase \cite{Jizba:2014taa}. The holographic two-point functions of four dimensional conformal gravity was computed in \cite{Ghodsi:2014hua}. It was shown that four dimensional conformal gravity plus a simple Neumann boundary condition can be used to get the semiclassical (or tree level) wavefunction of the universe of four dimensional asymptotically de-Sitter or Euclidean anti-de Sitter spacetimes \cite{Maldacena:2011mk}. A simple derivation of the equivalence between Einstein and Conformal Gravity (CG) with Neumann boundary conditions was provided in \cite{Anastasiou:2016jix}. It was argued that Weyl action should be added to the Einstein-Hilbert action \cite{tHooft:2016uxd}.

CG is also confronted with some challenges. It has been shown that CG does not agree with the predictions of general relativity in the limit of weak fields and slow motions, and it is therefore ruled out by Solar System observations \cite{Flanagan:2006ra}. It suggested that without dark matter CG can not explain the properties of X-ray galaxy clusters \cite{Diaferio:2008gh}. It is not able to describe the phenomenology of gravitational lensing \cite{Pireaux:2004id}. The cosmological models derived from CG are not likely to reproduce the observational properties of our Universe \cite{Elizondo:1994vh}.

In this paper, we will consider gravitational radiation in CG. We aim to find the equations of gravitational radiation and the energy-momentum tensor of the GWs. These results will be valuable for future observations of GWs to test gravity theories alternative to GR.

This paper is organised as follows. We begin with a review of the CG theory. In Section III, we consider GWs in CG. In Section IV, we will discuss the energy-momentum tensor of the GWs. Finally, we will briefly summarize and discuss our results.

\section{Basic equations for Conformal gravity}
Besides of the general coordinate invariance and equivalence principle structure of general relativity, CG possesses an additional local conformal symmetry in which the action is invariant under local conformal transformations on the metric: $g_{\mu\nu}\rightarrow e^{2\alpha(x)}g_{\mu\nu}$. This symmetry forbids the presence of any $\Lambda \sqrt{-g}{\rm d}^4x$ term in the action, so CG does not suffer from the cosmological constant problem \cite{Mannheim:2009qi}. Under such a symmetry, the action of CG in vacuum is given by
\begin{eqnarray}
\label{action}
\mathcal{I}&=&-\alpha_{\rm g} \int C_{\mu\nu\kappa\lambda} C^{\mu\nu\kappa\lambda}\sqrt{-g} {\rm d}^4x\\\nonumber
&=&-\alpha_{\rm g} \int\left[R_{\mu\nu\alpha\beta}R^{\mu\nu\alpha\beta}-2R_{\mu\nu}R^{\mu\nu}+\frac{1}{3}R^2\right]  \sqrt{-g} {\rm d}^4x\\\nonumber
&=&-2\alpha_{\rm g} \int\left[R_{\mu\nu}R^{\mu\nu}-\frac{1}{3}R^2\right]  \sqrt{-g} {\rm d}^4x,
\end{eqnarray}
where $C_{\mu\nu\kappa\lambda}$ the Weyl tensor and $\alpha_{\rm g}$ is a dimensionless coupling constant, unlike general relativity. To obtain the last equation, we have taken into account the fact that the Gauss-Bonnet term, $R_{\mu\nu\alpha\beta}R^{\mu\nu\alpha\beta}-4R_{\mu\nu}R^{\mu\nu}+R^2$, is a total derivative in four-dimension spacetime. Here we take the signature of the metric as $(-, +, +, +)$ and the Rieman tensor is defined as $R^{\lambda}_{~\mu\nu\kappa}=\partial_\kappa\Gamma^{\lambda}_{~\mu\nu}-\partial_\nu\Gamma^{\lambda}_{~\mu\kappa}+\Gamma^{\alpha}_{~\mu\nu}\Gamma^{\lambda}_{~\alpha\kappa}-\Gamma^{\alpha}_{~\mu\kappa}\Gamma^{\lambda}_{~\alpha\nu}$ from which the Ricci tensor is obtained $R_{\mu\nu}=g^{\alpha\beta}R_{\alpha\mu\beta\nu}$, and the Ricci scalar is defined as $R=g^{\mu\nu}R_{\mu\nu}$. If to include the matter in the action, the energy-momentum tensor of matter should also be Weyl invariant. Due to the conformal invariance one can choose a gauge in which the scalar field is constant. Then it is obvious that the Ricci scalar term belongs to the gravity part of the theory and cannot be discarded in the vacuum case. The only way this can be achieved is by choosing the vacuum expectation value of the scalar as zero. But this means that the fermion mass term also vanishes, which means that one can consider only massless matter with this theory. Hence, one should either say that one only consider massless matter or one has to take the Ricci term into account (for details see \cite{Mannheim1990}) in CG. The most general local matter action for a generic scalar and spinor field coupled conformally to gravity has been proposed in \cite{Mannheim1990}. Here we focus on the vacuum case corresponding to conformal gravity with massless matter. Variation with respect to the metric generates the field equations
\begin{eqnarray}
\label{field}
4\alpha_{\rm g}W_{\mu\nu}=0,
\end{eqnarray}
where
\begin{eqnarray}
\label{w}
\nonumber
W_{\mu\nu}=&-&\frac{1}{6}g_{\mu\nu}R^{;\lambda}_{~~;\lambda}+\frac{2}{3}R_{;\mu;\nu}+R_{\mu\nu;\lambda}^{~~~~~;\lambda}-R_{\lambda\nu;\mu}^{~~~~~;\lambda}-R_{\lambda\mu;\nu}^{~~~~~;\lambda}\\
&+&\frac{2}{3}RR_{\mu\nu}-2R_{\mu\lambda}R_{\nu}^{~\lambda}+\frac{1}{2}g_{\mu\nu}R_{\lambda\kappa}R^{\lambda\kappa}-\frac{1}{6}g_{\mu\nu}R^2,
\end{eqnarray}
Since $W^{\mu\nu}$ is obtain from an action that is both conformal invariant and general coordinate invariant, it is traceless and kinematically covariantly conserved: $W^{\mu}_{~\mu}\equiv g_{\mu\nu}W^{\mu\nu}=0$ and $W^{\mu\nu}_{~~~;\nu}=0$.

\section{Gravitational waves in Conformal gravity}
Here we are interested in vacuum GWs of CG. Recently the GWs of CG with matter are discussed in \cite{Caprini:2018oqe}, however, the results in vacuum case cannot be derived simply from the non-vacuum case by letting the energy-momentum tensor as zero. The linearized framework provides a natural way to study gravitational waves, which is a weak-field approximation that assumes small deviations from a flat background
\begin{eqnarray}
\label{per}
g_{\mu\nu}=\eta_{\mu\nu}+h_{\mu\nu},
\end{eqnarray}
where $|h_{\mu\nu}|\sim \epsilon$ which is a small parameter. We will consider terms up to $\mathcal{O}(\epsilon)$. Thus the inverse metric is $g^{\mu\nu}=\eta^{\mu\nu}-h^{\mu\nu}$ where the indices are raised by used the Minkowski metric. To the first-order, the covariant derivative of any perturbed quantity will be the same as the partial derivative, so the connection and the Riemann tensor are, respectively, given by
%%%%%%%%%%%%%%%%%
\begin{eqnarray}
\label{conect}
\Gamma^{(1)\rho}_{~~~~\mu\nu}=\frac{1}{2}\eta^{\rho\lambda}(\partial_{\mu}h_{\nu\lambda}+\partial_{\nu}h_{\mu\lambda}-\partial_{\lambda}h_{\mu\nu}),
\end{eqnarray}
%%%%%%%%%%%%%%%%%%%%%%
\begin{eqnarray}
\label{conect}
R^{(1)\lambda}_{~~~~\mu\nu\rho}=\frac{1}{2}(\partial_{\mu}\partial_{\rho}h^{\lambda}_{~\nu}+\partial^{\lambda}\partial_{\nu}h_{\mu\rho}-\partial_{\mu}\partial_{\nu}h^{\lambda}_{~\rho}-\partial^{\lambda}\partial_{\rho}h_{\mu\nu}).
\end{eqnarray}
Contracting the Riemann tensor gives the Ricci tensor
\begin{eqnarray}
\label{ricci}
R^{(1)}_{~~~\mu\nu}=\frac{1}{2}(\Box h_{\mu\nu}+\partial_{\mu}\partial_{\nu}h-\partial_{\mu}\partial_{\lambda}h^{\lambda}_{~\nu}-\partial_{\nu}\partial_{\lambda}h^{\lambda}_{~\mu}),
\end{eqnarray}
where the d'Alembertian operator is $\Box=\eta^{\mu\nu}\partial_{\mu}\partial_{\nu}$. Contracting the Ricci tensor gives the first-order Ricci scalar
\begin{eqnarray}
\label{riccisc}
R^{(1)}=\Box h-\partial_{\mu}\partial_{\nu}h^{\mu\nu}.
\end{eqnarray}
Inserting Eqs. (\ref{ricci}) and (\ref{riccisc}) into (\ref{w}) and retaining terms to the first-order, we obtain
\begin{eqnarray}
\label{w1}
W^{(1)}_{~~~\mu\nu}=-\frac{1}{6}\eta_{\mu\nu}\Box R^{(1)}+\frac{2}{3}R^{(1)}_{,\mu\nu}+\Box R^{(1)}_{\mu\nu}-R_{\lambda\nu,\mu}^{(1)~~,\lambda}-R_{\lambda\mu,\nu}^{(1)~~,\lambda}.
\end{eqnarray}

In general relativity, if we defines the trace-reversed perturbation $\bar{h}_{\mu\nu}=h_{\mu\nu}-\frac{1}{2}\eta_{\mu\nu}h$ and impose the Lorenz gauge $\partial^{\mu}\bar{h}_{\mu\nu}=0$, the linearized vacuum Einstein field equations reduce to the wave equation
\begin{eqnarray}
\label{e1}
\Box \bar{h}_{\mu\nu}=0.
\end{eqnarray}
We can apply this similar standard reasoning within the CG framework and find a quantity $\bar{h}_{\mu\nu}$ that satisfies a wave equation when linearizing the field equations (\ref{w}). The rank-two tensors in linearized conformal gravity are: $h_{\mu\nu}$, $\eta_{\mu\nu}$, $R^{(1)}_{\mu\nu}$, and $\partial_{\mu}\partial_{\nu}$. In order to eliminate $R^{(1)}_{\mu\nu}$, we will try the simper combination $\eta_{\mu\nu}R^{(1)}$. The linearized field equations (\ref{w}) is forth-order, we hope to get second-order wave equations which can be easily solved, so we look for a solution with the following form
\begin{eqnarray}
\label{h1}
\bar{h}_{\mu\nu}=\Box h_{\mu\nu}+\alpha\eta_{\mu\nu}\Box h+\beta\eta_{\mu\nu}R^{(1)},
\end{eqnarray}
where $\alpha$ and $\beta$ are constants. Taking the trace of Eq. (\ref{h1}) yields
\begin{eqnarray}
\label{ht}
\bar{h}=(4\alpha+1)\Box h+4\beta R^{(1)}.
\end{eqnarray}
So we can eliminate $h_{\mu\nu}$ in terms of $\bar{h}_{\mu\nu}$ to give
\begin{eqnarray}
\label{h11}
\Box h_{\mu\nu}=\bar{h}_{\mu\nu}-\frac{\alpha}{4\alpha+1}\eta_{\mu\nu}\bar{h}-\frac{\beta}{4\alpha+1}\eta_{\mu\nu}R^{(1)},
\end{eqnarray}
and
\begin{eqnarray}
\label{ht1}
\Box h=\frac{1}{4\alpha+1}\bar{h}- \frac{4\beta}{4\alpha+1}R^{(1)}.
\end{eqnarray}
Inserting Eqs. (\ref{h11}) and (\ref{ht1}) into $\Box R^{(1)}_{\mu\nu}$ yields
\begin{eqnarray}
\label{bricci}
\Box R^{(1)}_{\mu\nu}=\frac{1}{2}\left[\Box \bar{h}_{\mu\nu}-\frac{\alpha\eta_{\mu\nu}}{4\alpha+1}\Box \bar{h}-\partial_{\mu}\partial^{\lambda}\bar{h}_{\lambda\nu}-\partial_{\nu}\partial^{\lambda}\bar{h}_{\lambda\mu}
+\frac{2\alpha+1}{4\alpha+1}\partial_{\mu}\partial_{\nu}\bar{h}-\frac{2\beta}{4\alpha+1}\partial_{\mu}\partial_{\nu}R^{(1)}-\frac{\beta\eta_{\mu\nu}}{4\alpha+1}\Box R^{(1)}\right].
\end{eqnarray}
Using Eq. (\ref{bricci}), $W^{(1)}_{\mu\nu}$ can be rewritten as
\begin{eqnarray}
\label{bw1}
W^{(1)}_{\mu\nu}=\frac{1}{2}\left[\Box \bar{h}_{\mu\nu}-\frac{\alpha\eta_{\mu\nu}}{4\alpha+1}\Box \bar{h}-\partial_{\mu}\partial^{\lambda}\bar{h}_{\lambda\nu}-\partial_{\nu}\partial^{\lambda}\bar{h}_{\lambda\mu}
+\frac{2\alpha+1}{4\alpha+1}\partial_{\mu}\partial_{\nu}\bar{h}- \frac{3\beta+4\alpha+1}{3(4\alpha+1)}\left(6\partial_{\mu}\partial_{\nu}R^{(1)}+\eta_{\mu\nu}\Box R^{(1)}\right)\right],
\end{eqnarray}
where we have used $R^{(1)~~,\lambda}_{\lambda\nu,\mu}=R^{(1)~~,\lambda}_{\lambda\mu,\nu}=\frac{1}{2}\partial_{\mu}\partial_{\nu}R^{(1)}$. The third and fourth term can be eliminated if Lorentz gauge conditions are taking. This can be done via coordinate transformation. Considering an infinitesimal coordinate transformation,
$x^{\mu}\rightarrow x'^{\mu}=x^\mu+\xi^\mu$, we get
\begin{eqnarray}
&&h'_{\mu\nu}=h_{\mu\nu}-\partial_{\mu}\xi_\nu-\partial_{\nu}\xi_\mu,\\
&&h'=h-2\partial_\sigma\xi^\sigma,\\
&&\bar{h}'_{\mu\nu}=\Box h'_{\mu\nu}+\alpha\eta_{\mu\nu}\Box h'+\beta\eta_{\mu\nu}R'^{(1)}=\bar{h}_{\mu\nu}-\Box (\partial_{\mu}\xi_\nu+\partial_{\nu}\xi_\mu)+\eta_{\mu\nu}\Box\partial_\sigma\xi^\sigma,\\
\label{hbc}
&&\bar{h}'=\bar{h}+2\Box\partial_\sigma\xi^\sigma,
\end{eqnarray}
where the index was raised or lowered by using the Minkowski metric $\eta_{\mu\nu}$. If choosing $\xi_\mu$ so that it satisfies the equation
\begin{eqnarray}
\Box^2\xi_\nu=\partial^{\mu}\bar{h}_{\mu\nu},
\end{eqnarray}
then we have the Lorentz gauge condition $\partial^{\mu}\bar{h}'_{\mu\nu}=0$. Using this gauge conditions, the third and fourth term in (\ref{bw1}) are zero. To obtain the wave equation, the fifth and the last term should also be set to zero. These two terms can not be eliminated by choosing gauge conditions. We can, however, choose freely the parameters $\alpha$ and $\beta$ in the equation (\ref{h1}) to eliminate these two terms, and this can be done by setting the coefficients of these two terms to zero. It is easily to find when taking $\alpha=-\frac{1}{2}$ and $\beta=\frac{1}{3}$, the fifth and the last term can be eliminated. Then we obtain the first-order linear vacuum equations (\ref{field}) as
\begin{eqnarray}
\label{cw1}
\frac{1}{2}\Box \bar{h}_{\mu\nu}-\frac{1}{4}\eta_{\mu\nu}\Box\bar{h}=0.
\end{eqnarray}
Taking trace and yields
\begin{eqnarray}
\label{cw2}
\frac{1}{2}\Box \bar{h}-\Box\bar{h}=0,
\end{eqnarray}
This equation leads to
\begin{eqnarray}
\label{cws}
\Box \bar{h}=0.
\end{eqnarray}
Combining Eqs. (\ref{cw1}) and (\ref{cws}), we finally obtain
\begin{eqnarray}
\label{gw}
\Box \bar{h}_{\mu\nu}=0,
\end{eqnarray}
which seemingly has the same form of the GWs has in general relativity, they are different from each other in fact: with different $\bar{h}_{\mu\nu}$ and with different energy-momentum tensor as shown below. The solution takes the form
\begin{eqnarray}
\label{gws}
\bar{h}_{\mu\nu}=q_{\mu\nu}\exp{(ik_{\lambda}x^{\lambda})}.
\end{eqnarray}
where $k_\lambda$ is a four-wavevector and satisfies $\eta_{\mu\nu}k^{\mu}k^{\nu}=0$ and $k^{\mu}q_{\mu\nu}=0$. For a wave traveling along the $z$-axis, $k^{\mu}=\omega(1,0,0,1)$ with $\omega$ the angular frequency. The Lorentz gauge condition can not fixed the gauge freedom completely, it leaves a residual coordinate transformation with $\Box^2\xi_\nu=0$. If $\xi_\mu$ also satisfies the equation $\bar{h}=-2\Box\partial_\sigma\xi^\sigma$, we then have $\bar{h}'=0$. We may impose further constrains upon $q_{\mu\nu}$: $q_{0\nu}=0$, and can define
\begin{equation}
\label{sw}
\left[q_{\mu\nu}\right] =
\begin{bmatrix}
0 & 0 & 0 & 0\\
0 & q_+ & q_\times & 0\\
0 & q_\times & -q_+ & 0\\
0 & 0 & 0 & 0
\end{bmatrix},
\end{equation}
where $q_+$ and $q_\times$ are constants representing the amplitudes of the two transverse polarizations of gravitational radiation. Here we note that equations (\ref{gws}) and (\ref{sw}) may mislead one to conclude that there are only two independent components. The degree of freedom of the linearized conformal gravity was discussed in detail in \cite{RIEGERT1984110}, which was consistent with the results obtained by using the canonical analysis \cite{Nieuwenhuizen1982}. In \cite{RIEGERT1984110}, Riegert introduced $\bar{h}_{\mu\nu}=h_{\mu\nu}-\frac{1}{4}\eta_{\mu\nu}h$ and obtain fourth-order wave equations $\Box^2\bar{h}_{\mu\nu}=0$ under conformal gauge conditions $V_\mu=\frac{1}{3}\bar{h}_{\alpha\beta,\mu}^{~~~~,\alpha,\beta}=0$. Riegert has shown that the plane wave solutions of linearized conformal gravity propagate six physical degrees of freedom, corresponding to massless spin-2 and spin-1 ordinary particles and a massless spin-2 ghost particle. Here we introduce the variable of field (\ref{h1}) and the Lorentz gauge conditions $\partial^{\mu}\bar{h}_{\mu\nu}=0$ and obtain $\Box\bar{h}_{\mu\nu}=0$ which is second-order equations differentiating from the fourth-order wave equations obtained in \cite{RIEGERT1984110}, this is because the variable of field (\ref{h1}) introduced here is proportional to the second derivative of the metric perturbation. Though if we insert $\bar{h}_{\mu\nu}$ (as functions of $h_{\mu\nu}$) into the wave equations, both method lead to the same equations $\Box^2 h_{\mu\nu}-\frac{1}{4}\eta_{\mu\nu}\Box^2 h=0$ but with gauge conditions $\partial_{\mu}\partial^{\alpha}\partial^{\beta}h_{\alpha\beta}=3\Box\partial^{\alpha}h_{\alpha\mu}$ in \cite{RIEGERT1984110} and $\partial_{\mu}\partial^{\alpha}\partial^{\beta}h_{\alpha\beta}=\Box\partial^{\alpha}h_{\alpha\mu}$ here, it is $h_{\mu\nu}$, not the second derivative of the metric, that is the actual perturbation. To obtain the solution for the original metric perturbation from solution (\ref{gws}) for the new metric perturbation, one should to assume the form of the solution for the original metric perturbation as done in \cite{RIEGERT1984110}, then insert it into the equation (\ref{h1}) for $\bar{h}_{\mu\nu}$, comparing with the solution (\ref{gws}), one can obtain the solution for the original metric perturbation. The particle content of linearized conformal gravity should be investigated in the forth-order wave equations, see \cite{RIEGERT1984110,Caprini:2018oqe} for details.

\section{Energy-Momentum tensor of the gravitational waves}
Physically, we could expect the gravitational field to carry energy-momentum, but as it is well known that it is difficult to define an energy-momentum tensor for a gravitational field. Nevertheless, one can regard the
linearised theory as describing a simple rank-2 tensor field $h_{\mu\nu}$ in Cartesian inertial coordinates propagating in a fixed Minkowski spacetime background, and then can assign an energy-momentum tensor to this field in Minkowski spacetime. As was discussed above, the linearised gravitational theory ignores the energy-momentum associated with the gravitational field itself. To include this contribution, and thereby go beyond the linearised theory, we must modify the linearised field equations to read
\begin{eqnarray}
W^{(1)}_{\mu\nu}=\frac{1}{4\alpha_{\rm g}}t_{\mu\nu},
\end{eqnarray}
where $t_{\mu\nu}$ is the energy-momentum tensor of the gravitational
field itself. Rearranging this equation gives
\begin{eqnarray}
W^{(1)}_{\mu\nu}-\frac{1}{4\alpha_{\rm g}}t_{\mu\nu}=0.
\end{eqnarray}
Returning to the equations (\ref{w}), we can expand beyond first order to obtain
\begin{eqnarray}
W_{\mu\nu}\equiv W^{(1)}_{\mu\nu}+W^{(2)}_{\mu\nu}+...=0.
\end{eqnarray}
where indexes $(i)$ indicate the order of the expansion in $h_{\mu\nu}$. To a good approximation (up to $\mathcal{O}(\epsilon^3)$), we should make the identification
\begin{eqnarray}
\label{wenm}
t_{\mu\nu}=-4\alpha_{\rm g}W^{(2)}_{\mu\nu}.
\end{eqnarray}
Since the energy-momentum of a gravitational field at a point in spacetime has no real meaning, this suggests that in order to probe the physical curvature of the spacetime one should average $W^{(2)}_{\mu\nu}$ over a small region at each point in spacetime, which gives a gauge-invariant measure of the gravitationa field strength of the GWs themselves. The average is carried out over a region of length scale $d$ with $\lambda\ll d\ll L$, where $\lambda$ is the GW wavelength and $L$ is the background radius. The average of a tensor is defined as \cite{Isaacson:1967zz, Isaacson:1968zza}
\begin{eqnarray}
\label{aver}
\langle A_{\mu\nu}\rangle=\int j^{\alpha}_{\mu}(x, x')j^{\beta}_{\nu}(x, x')A_{\alpha\beta}(x')f(x, x')\sqrt{-\bar{g}(x')}{\rm d}^4x'.
\end{eqnarray}
where $\langle\cdot\cdot\cdot\rangle$ denotes the averaging process, $\bar{g}_{\mu\nu}$ is the metric of the background, $j^{\alpha}_{\mu}$ is the bivector of geodesic parallel displacement, and $f(x, x')$ is a weight function that falls quickly and smoothly to zero when $\mid x-x'\mid>d$, such that $\int_{\rm all space}f(x, x')\sqrt{-\bar{g}(x')}{\rm d}^4x'=1$. Therefore we should replace Eq. (\ref{wenm}) by
\begin{eqnarray}
\label{wenm1}
t_{\mu\nu}=-4\alpha_{\rm g}\big{\langle} W^{(2)}_{\mu\nu}\big{\rangle}.
\end{eqnarray}
To calculate the energy-momentum of GWs, we must expand each team in $W_{\mu\nu}$ up to the second order in $h_{\mu\nu}$. The second-order Ricci tensor is
\begin{eqnarray}
R^{(2)}_{\mu\nu}=\partial_{\nu}\Gamma^{(2)\sigma}_{~~~~\mu\sigma}-
\partial_{\sigma}\Gamma^{(2)\sigma}_{~~~~\mu\nu}+
\Gamma^{(1)\rho}_{~~~~\mu\sigma}\Gamma^{(1)\sigma}_{~~~~\rho\nu}-
\Gamma^{(1)\rho}_{~~~~\mu\nu}\Gamma^{(1)\sigma}_{~~~~\rho\sigma},
\end{eqnarray}
where $\Gamma^{(1)\sigma}_{~~~~\mu\nu}$ is presented in equation (\ref{conect}) and $\Gamma^{(2)\sigma}_{~~~~\mu\nu}$ is given by
\begin{eqnarray}
\Gamma^{(2)\sigma}_{~~~~\mu\nu}=-\frac{1}{2}h^{\sigma\lambda}(\partial_{\mu}h_{\nu\lambda}+\partial_{\nu}h_{\mu\lambda}
-\partial_{\lambda}h_{\mu\nu})
\end{eqnarray}
Since we average over all directions at each point, first derivatives average
to zero: $\langle\partial_{\mu}f\rangle=0$ for any function of position $f$. This has the important consequence that
$\langle\partial_{\mu}fg\rangle=-\langle\partial_{\mu}gf\rangle$ \cite{Saffer:2017ywl,CambridgeGeneral}. Repeated application of this, we obtain the terms in $W^{(2)}_{\mu\nu}$ as follow, respectively
\begin{eqnarray}
\langle(g_{\mu\nu}\nabla^{\alpha}\nabla_{\alpha}R)^{(2)}\rangle=
\langle\eta_{\mu\nu}\eta^{\alpha\beta}\partial_{\lambda}\Gamma^{(1)\lambda}_{~~~~\alpha\beta}R^{(1)}+2\eta_{\mu\nu}\partial^{\alpha}\Gamma^{(1)\lambda}_{~~~~\alpha\rho}
R^{(1)\rho}_{~~\lambda}-\eta_{\mu\nu}h^{\alpha\beta}\partial_{\alpha\beta}R^{(1)}-\eta_{\mu\nu}h^{\rho\sigma}\Box R^{(1)}_{\rho\sigma}+h_{\mu\nu}\Box R^{(1)}\rangle,
\end{eqnarray}
%%%%%%%%%%%%%%%%
\begin{eqnarray}
\langle(\nabla^{\alpha}\nabla_{\alpha}R_{\mu\nu})^{(2)}\rangle=
\langle\eta^{\alpha\beta}\partial_{\lambda}\Gamma^{(1)\lambda}_{~~~~\alpha\beta}R^{(1)}_{\mu\nu}+\partial^{\alpha}\Gamma^{(1)\lambda}_{~~~~\alpha\mu}
R^{(1)}_{\lambda\nu}+\partial^{\alpha}\Gamma^{(1)\lambda}_{~~~~\alpha\nu}
R^{(1)}_{\lambda\mu}-h^{\alpha\beta}\partial_{\alpha\beta}R^{(1)}_{\mu\nu}\rangle,
\end{eqnarray}
%%%%%%%%%%%%%%%%
\begin{eqnarray}
\langle(\nabla^{\alpha}\nabla_{\mu}R_{\nu\alpha})^{(2)}\rangle=
\langle-h^{\alpha\beta}\partial_{\alpha}\partial_{\mu}R^{(1)}_{\nu\beta}
+\partial_{\lambda}\Gamma^{(1)\lambda}_{~~~~\alpha\mu}
R^{(1)\alpha}_{~~\nu}+\partial_{\mu}\Gamma^{(1)\lambda}_{~~~~\alpha\nu}
R^{(1)\alpha}_{~~\lambda}+\eta^{\alpha\beta}\partial_{\mu}\Gamma^{(1)\lambda}_{~~~~\alpha\beta}
R^{(1)}_{\nu\lambda}\rangle,
\end{eqnarray}
%%%%%%%%%%%%%%%%%
\begin{eqnarray}
\langle(\nabla_{\nu}\nabla_{\mu}R)^{(2)}\rangle=
\langle-h^{\alpha\beta}\partial_{\mu}\partial_{\nu}R^{(1)}_{\alpha\beta}
+\partial_{\lambda}\Gamma^{(1)\lambda}_{~~~~\mu\nu}
R^{(1)}+2\partial_{\mu}\Gamma^{(1)\lambda}_{~~~~\nu\sigma}
R^{(1)\sigma}_{~~\lambda}\rangle,
\end{eqnarray}
%%%%%%%%%%%%%%%%%%
\begin{eqnarray}
\langle(R_{\mu\alpha}R^{~\alpha}_{\nu})^{(2)}\rangle=
\langle\eta^{\alpha\beta}R^{(1)}_{\mu\alpha}R^{(1)}_{\nu\beta}\rangle,
\end{eqnarray}
%%%%%%%%%%%%%%%
\begin{eqnarray}
\langle(g_{\mu\nu}R_{\alpha\beta}R^{\alpha\beta})^{(2)}\rangle=
\langle \eta_{\mu\nu}\eta^{\alpha\rho}\eta^{\beta\sigma}R^{(1)}_{\alpha\beta}R^{(1)}_{\rho\sigma}\rangle,
\end{eqnarray}
%%%%%%%%%%%%%%%%%%%
\begin{eqnarray}
\langle(RR_{\mu\nu})^{(2)}\rangle=
\langle R^{(1)}R^{(1)}_{\mu\nu}\rangle,
\end{eqnarray}
%%%%%%%%%%%%%%%%%%%%
\begin{eqnarray}
\langle(g_{\mu\nu}R^2)^{(2)}\rangle=
\langle \eta_{\mu\nu}R^{(1)2}\rangle,
\end{eqnarray}
Inserting the $R^{(1)}$, $R^{(1)}_{\mu\nu}$, and $\Gamma^{(1)\lambda}_{~~~~\mu\nu}$ into these equations, and using the wave equations (\ref{gw}), the vanishing trace $\bar{h}=0$, and the Lorentz gauge $\partial^{\mu}\bar{h}_{\mu\nu}=0$, we obtain after a rather cumbersome calculation
\begin{eqnarray}
t_{\mu\nu}&=&-4\alpha_{\rm g}\big{\langle} W^{(2)}_{\mu\nu}\big{\rangle}\\
&=&-4\alpha_{\rm g}\Big
{\langle}\Big[\frac{1}{2}\partial_{\mu}\partial_{\nu}h^{\alpha\beta}\partial_{\alpha}\partial^{\lambda}h_{\lambda\beta}+\frac{3}{32}\Box h \partial_{\mu}\partial_{\nu}h-\frac{1}{2}\partial_{\mu}\partial_{\nu}h^{\alpha\beta}\Box h_{\alpha\beta}\Big]\Big{\rangle}\\
&=&2\alpha_{\rm g}\Big
{\langle}\bar{h}_{\alpha\beta}\frac{\partial_{\mu}\partial_{\nu}}{\Box}\bar{h}^{\alpha\beta}\Big{\rangle}.
\end{eqnarray}
where $1/\Box=1/k^2$ in momentum space. It is obvious that this energy-momentum tensor is symmetric. Using the vanishing trace, the Lorentz gauge, and $W^{(1)}_{\mu\nu}=0$ (implying that $\Box R^{(1)}_{\mu\nu}=\frac{1}{6}\eta_{\mu\nu}\Box R^{(1)}+\frac{1}{3}\partial_{\mu}\partial_{\nu}R^{(1)}$), it can also been shown that this energy-momentum tensor is traceless. Because of the operator $1/\Box$, the energy carried by GWs in vacuum case diverges in momentum space, this due to the massless ghost contained in CG.

\section{Conclusions and discussions}
We have discussed the gravitational radiation in conformal gravity theory. We have linearized the field equations and obtain the second-order wave equation after introducing the suitable transformation for $h_{\mu\nu}$. We have also derived the effective energy-momentum tensor for the gravitational radiation. Unfortunately, however, the energy carried by GWs in vacuum case diverge in momentum space. The methods presented here can be applied to investigate other alternative gravity theories.

\begin{acknowledgments}
We thank Doctors Yungui Gong, Liu Zhao, and Bin Hu for useful discussions. This study is supported in part by National Natural Science Foundation of China (Grant Nos. 11273010 and 11147028), Hebei Provincial Natural Science Foundation of China (Grant No. A2014201068), the Outstanding Youth Fund of Hebei University (No. 2012JQ02), and the Midwest universities comprehensive strength promotion project.
\end{acknowledgments}

\bibliographystyle{elsarticle-num}%{ieeetr}%
\bibliography{ref}

\begin{thebibliography}{10}
\expandafter\ifx\csname url\endcsname\relax
  \def\url#1{\texttt{#1}}\fi
\expandafter\ifx\csname urlprefix\endcsname\relax\def\urlprefix{URL }\fi
\expandafter\ifx\csname href\endcsname\relax
  \def\href#1#2{#2} \def\path#1{#1}\fi

\bibitem{Abbott:2016blz}
B.~P. Abbott, et~al., {Observation of Gravitational Waves from a Binary Black
  Hole Merger}, Phys. Rev. Lett. 116~(6) (2016) 061102.
\newblock \href {http://arxiv.org/abs/1602.03837} {\path{arXiv:1602.03837}},
  \href {http://dx.doi.org/10.1103/PhysRevLett.116.061102}
  {\path{doi:10.1103/PhysRevLett.116.061102}}.

\bibitem{Abbott:2017oio}
B.~P. Abbott, et~al., {GW170814: A Three-Detector Observation of Gravitational
  Waves from a Binary Black Hole Coalescence}, Phys. Rev. Lett. 119~(14) (2017)
  141101.
\newblock \href {http://arxiv.org/abs/1709.09660} {\path{arXiv:1709.09660}},
  \href {http://dx.doi.org/10.1103/PhysRevLett.119.141101}
  {\path{doi:10.1103/PhysRevLett.119.141101}}.

\bibitem{GBM:2017lvd}
B.~P. Abbott, et~al., {Multi-messenger Observations of a Binary Neutron Star
  Merger}, Astrophys. J. 848~(2) (2017) L12.
\newblock \href {http://arxiv.org/abs/1710.05833} {\path{arXiv:1710.05833}},
  \href {http://dx.doi.org/10.3847/2041-8213/aa91c9}
  {\path{doi:10.3847/2041-8213/aa91c9}}.

\bibitem{Li:2017iup}
T.~Li, et~al., {Insight-HXMT observations of the first binary neutron star
  merger GW170817}, Sci. China Phys. Mech. Astron. 61 (2018) 031011.
\newblock \href {http://arxiv.org/abs/1710.06065} {\path{arXiv:1710.06065}},
  \href {http://dx.doi.org/10.1007/s11433-017-9107-5}
  {\path{doi:10.1007/s11433-017-9107-5}}.

\bibitem{AmaroSeoane:2012km}
P.~Amaro-Seoane, et~al., {eLISA/NGO: Astrophysics and cosmology in the
  gravitational-wave millihertz regime}, GW Notes 6 (2013) 4--110.
\newblock \href {http://arxiv.org/abs/1201.3621} {\path{arXiv:1201.3621}}.

\bibitem{Yang:2011cp}
L.~Yang, C.-C. Lee, C.-Q. Geng, {Gravitational Waves in Viable $f(R)$ Models},
  JCAP 1108 (2011) 029.
\newblock \href {http://arxiv.org/abs/1106.5582} {\path{arXiv:1106.5582}},
  \href {http://dx.doi.org/10.1088/1475-7516/2011/08/029}
  {\path{doi:10.1088/1475-7516/2011/08/029}}.

\bibitem{Liang:2017ahj}
D.~Liang, Y.~Gong, S.~Hou, Y.~Liu, {Polarizations of gravitational waves in
  $f(R)$ gravity}, Phys. Rev. D95~(10) (2017) 104034.
\newblock \href {http://arxiv.org/abs/1701.05998} {\path{arXiv:1701.05998}},
  \href {http://dx.doi.org/10.1103/PhysRevD.95.104034}
  {\path{doi:10.1103/PhysRevD.95.104034}}.

\bibitem{Rizwana:2016qdq}
H.~Rizwana~Kausar, L.~Philippoz, P.~Jetzer, {Gravitational Wave Polarization
  Modes in $f(R)$ Theories}, Phys. Rev. D93~(12) (2016) 124071.
\newblock \href {http://arxiv.org/abs/1606.07000} {\path{arXiv:1606.07000}},
  \href {http://dx.doi.org/10.1103/PhysRevD.93.124071}
  {\path{doi:10.1103/PhysRevD.93.124071}}.

\bibitem{Capozziello:2008rq}
S.~Capozziello, C.~Corda, M.~F. De~Laurentis, {Massive gravitational waves from
  f(R) theories of gravity: Potential detection with LISA}, Phys. Lett. B669
  (2008) 255--259.
\newblock \href {http://arxiv.org/abs/0812.2272} {\path{arXiv:0812.2272}},
  \href {http://dx.doi.org/10.1016/j.physletb.2008.10.001}
  {\path{doi:10.1016/j.physletb.2008.10.001}}.

\bibitem{Corda:2007nr}
C.~Corda, {Massive gravitational waves from the $R^2$ theory of gravity:
  Production and response of interferometers}, Int. J. Mod. Phys. A23 (2008)
  1521--1535.
\newblock \href {http://arxiv.org/abs/0711.4917} {\path{arXiv:0711.4917}},
  \href {http://dx.doi.org/10.1142/S0217751X08038603}
  {\path{doi:10.1142/S0217751X08038603}}.

\bibitem{Corda:2007hi}
C.~Corda, {The production of matter from curvature in a particular linearized
  high order theory of gravity and the longitudinal response function of
  interferometers}, JCAP 0704 (2007) 009.
\newblock \href {http://arxiv.org/abs/astro-ph/0703644}
  {\path{arXiv:astro-ph/0703644}}, \href
  {http://dx.doi.org/10.1088/1475-7516/2007/04/009}
  {\path{doi:10.1088/1475-7516/2007/04/009}}.

\bibitem{Corda:2010zza}
C.~Corda, {Massive relic gravitational waves from f(R) theories of gravity:
  Production and potential detection}, Eur. Phys. J. C65 (2010) 257--267.
\newblock \href {http://arxiv.org/abs/1007.4077} {\path{arXiv:1007.4077}},
  \href {http://dx.doi.org/10.1140/epjc/s10052-009-1100-5}
  {\path{doi:10.1140/epjc/s10052-009-1100-5}}.

\bibitem{Alves:2009eg}
M.~E.~S. Alves, O.~D. Miranda, J.~C.~N. de~Araujo, {Probing the $f(R)$
  formalism through gravitational wave polarizations}, Phys. Lett. B679 (2009)
  401--406.
\newblock \href {http://arxiv.org/abs/0908.0861} {\path{arXiv:0908.0861}},
  \href {http://dx.doi.org/10.1016/j.physletb.2009.08.005}
  {\path{doi:10.1016/j.physletb.2009.08.005}}.

\bibitem{Alves:2010ms}
M.~E.~S. Alves, O.~D. Miranda, J.~C.~N. de~Araujo, {Extra polarization states
  of cosmological gravitational waves in alternative theories of gravity},
  Class. Quant. Grav. 27 (2010) 145010.
\newblock \href {http://arxiv.org/abs/1004.5580} {\path{arXiv:1004.5580}},
  \href {http://dx.doi.org/10.1088/0264-9381/27/14/145010}
  {\path{doi:10.1088/0264-9381/27/14/145010}}.

\bibitem{Berry:2011pb}
C.~P.~L. Berry, J.~R. Gair, {Linearized f(R) Gravity: Gravitational Radiation
  and Solar System Tests}, Phys. Rev. D83 (2011) 104022, [Erratum: Phys.
  Rev.D85,089906(2012)].
\newblock \href {http://arxiv.org/abs/1104.0819} {\path{arXiv:1104.0819}},
  \href {http://dx.doi.org/10.1103/PhysRevD.85.089906,
  10.1103/PhysRevD.83.104022} {\path{doi:10.1103/PhysRevD.85.089906,
  10.1103/PhysRevD.83.104022}}.

\bibitem{Alves:2016iks}
M.~Alves, P.~Moraes, J.~de~Araujo, M.~Malheiro, {Gravitational waves in
  $f(R,T)$ and $f(R,T^\phi)$ theories of gravity}, Phys. Rev. D94~(2) (2016)
  024032.
\newblock \href {http://arxiv.org/abs/1604.03874} {\path{arXiv:1604.03874}},
  \href {http://dx.doi.org/10.1103/PhysRevD.94.024032}
  {\path{doi:10.1103/PhysRevD.94.024032}}.

\bibitem{Kausar:2017uib}
H.~R. Kausar, {Polarization states of gravitational waves in modified
  theories}, Int. J. Mod. Phys. D26~(5) (2017) 1741010.
\newblock \href {http://dx.doi.org/10.1142/S0218271817410103}
  {\path{doi:10.1142/S0218271817410103}}.

\bibitem{Capozziello:2006ra}
S.~Capozziello, C.~Corda, {Scalar gravitational waves from scalar-tensor
  gravity: Production and response of interferometers}, Int. J. Mod. Phys. D15
  (2006) 1119--1150.
\newblock \href {http://dx.doi.org/10.1142/S0218271806008814}
  {\path{doi:10.1142/S0218271806008814}}.

\bibitem{Hou:2017bqj}
S.~Hou, Y.~Gong, Y.~Liu, {The Polarizations of Gravitational Waves in
  Scalar-Tensor Theory}\href {http://arxiv.org/abs/1704.01899}
  {\path{arXiv:1704.01899}}.

\bibitem{Zhang:2017srh}
X.~Zhang, T.~Liu, W.~Zhao, {Gravitational radiation from compact binary systems
  in screened modified gravity}, Phys. Rev. D95~(10) (2017) 104027.
\newblock \href {http://arxiv.org/abs/1702.08752} {\path{arXiv:1702.08752}},
  \href {http://dx.doi.org/10.1103/PhysRevD.95.104027}
  {\path{doi:10.1103/PhysRevD.95.104027}}.

\bibitem{Bamba:2013ooa}
K.~Bamba, S.~Capozziello, M.~De~Laurentis, S.~Nojiri, D.~S¨¢ez-G¨®mez, {No
  further gravitational wave modes in $F(T)$ gravity}, Phys. Lett. B727 (2013)
  194--198.
\newblock \href {http://arxiv.org/abs/1309.2698} {\path{arXiv:1309.2698}},
  \href {http://dx.doi.org/10.1016/j.physletb.2013.10.022}
  {\path{doi:10.1016/j.physletb.2013.10.022}}.

\bibitem{Capozziello:2015nga}
S.~Capozziello, A.~Stabile, {Gravitational waves in fourth order gravity},
  Astrophys. Space Sci. 358~(2) (2015) 27.
\newblock \href {http://dx.doi.org/10.1007/s10509-015-2425-1}
  {\path{doi:10.1007/s10509-015-2425-1}}.

\bibitem{Chakraborty:2017qve}
S.~Chakraborty, K.~Chakravarti, S.~Bose, S.~SenGupta, {Signatures of extra
  dimensions in gravitational waves from black hole quasi-normal modes}\href
  {http://arxiv.org/abs/1710.05188} {\path{arXiv:1710.05188}}.

\bibitem{Bueno:2016ypa}
P.~Bueno, P.~A. Cano, V.~S. Min, M.~R. Visser, {Aspects of general higher-order
  gravities}, Phys. Rev. D95~(4) (2017) 044010.
\newblock \href {http://arxiv.org/abs/1610.08519} {\path{arXiv:1610.08519}},
  \href {http://dx.doi.org/10.1103/PhysRevD.95.044010}
  {\path{doi:10.1103/PhysRevD.95.044010}}.

\bibitem{Weyl1918Reine}
H.~Weyl, Reine infinitesimalgeometrie, Mathematische Zeitschrift 2~(3-4) (1918)
  384--411.

\bibitem{Scholz:2011za}
E.~Scholz, {Weyl geometry in late 20th century physics}\href
  {http://arxiv.org/abs/1111.3220} {\path{arXiv:1111.3220}}.

\bibitem{Mannheim:2005bfa}
P.~D. Mannheim, {Alternatives to dark matter and dark energy}, Prog. Part.
  Nucl. Phys. 56 (2006) 340--445.
\newblock \href {http://arxiv.org/abs/astro-ph/0505266}
  {\path{arXiv:astro-ph/0505266}}, \href
  {http://dx.doi.org/10.1016/j.ppnp.2005.08.001}
  {\path{doi:10.1016/j.ppnp.2005.08.001}}.

\bibitem{Yang:2013skk}
R.~Yang, B.~Chen, H.~Zhao, J.~Li, Y.~Liu, {Test of conformal gravity with
  astrophysical observations}, Phys. Lett. B727 (2013) 43--47.
\newblock \href {http://arxiv.org/abs/1311.2800} {\path{arXiv:1311.2800}},
  \href {http://dx.doi.org/10.1016/j.physletb.2013.10.035}
  {\path{doi:10.1016/j.physletb.2013.10.035}}.

\bibitem{OBrien:2011vks}
J.~G. O'Brien, P.~D. Mannheim, {Fitting dwarf galaxy rotation curves with
  conformal gravity}, Mon. Not. Roy. Astron. Soc. 421 (2012) 1273.
\newblock \href {http://arxiv.org/abs/1107.5229} {\path{arXiv:1107.5229}},
  \href {http://dx.doi.org/10.1111/j.1365-2966.2011.20386.x}
  {\path{doi:10.1111/j.1365-2966.2011.20386.x}}.

\bibitem{Mannheim:2011is}
P.~D. Mannheim, {Cosmological Perturbations in Conformal Gravity}, Phys. Rev.
  D85 (2012) 124008.
\newblock \href {http://arxiv.org/abs/1109.4119} {\path{arXiv:1109.4119}},
  \href {http://dx.doi.org/10.1103/PhysRevD.85.124008}
  {\path{doi:10.1103/PhysRevD.85.124008}}.

\bibitem{Amarasinghe:2018jkv}
A.~Amarasinghe, M.~G. Phelps, P.~D. Mannheim, {Cosmological perturbations in
  conformal gravity II}\href {http://arxiv.org/abs/1805.06807}
  {\path{arXiv:1805.06807}}.

\bibitem{RIEGERT1984110}
R.~J. Riegert,
  \href{http://www.sciencedirect.com/science/article/pii/0375960184906480}{The
  particle content of linearized conformal gravity}, Physics Letters A 105~(3)
  (1984) 110 -- 112.
\newblock \href
  {http://dx.doi.org/https://doi.org/10.1016/0375-9601(84)90648-0}
  {\path{doi:https://doi.org/10.1016/0375-9601(84)90648-0}}.
\newline\urlprefix\url{http://www.sciencedirect.com/science/article/pii/0375960184906480}

\bibitem{Diaferio:2011kc}
A.~Diaferio, L.~Ostorero, V.~F. Cardone, {Gamma-ray bursts as cosmological
  probes: LambdaCDM vs. conformal gravity}, JCAP 1110 (2011) 008.
\newblock \href {http://arxiv.org/abs/1103.5501} {\path{arXiv:1103.5501}},
  \href {http://dx.doi.org/10.1088/1475-7516/2011/10/008}
  {\path{doi:10.1088/1475-7516/2011/10/008}}.

\bibitem{Varieschi:2014pca}
G.~U. Varieschi, {Astrophysical Tests of Kinematical Conformal Cosmology in
  Fourth-Order Conformal Weyl Gravity}, Galaxies 2~(4) (2014) 577--600.
\newblock \href {http://arxiv.org/abs/1410.2944} {\path{arXiv:1410.2944}},
  \href {http://dx.doi.org/10.3390/galaxies2040577}
  {\path{doi:10.3390/galaxies2040577}}.

\bibitem{Roberts:2017nkm}
C.~Roberts, K.~Horne, A.~O. Hodson, A.~D. Leggat, {Tests of $\Lambda$CDM and
  Conformal Gravity using GRB and Quasars as Standard Candles out to $z \sim
  8$}\href {http://arxiv.org/abs/1711.10369} {\path{arXiv:1711.10369}}.

\bibitem{Zhang:2017amt}
H.~Zhang, Y.~Zhang, X.-Z. Li, {Dynamical spacetimes in conformal gravity},
  Nucl. Phys. B921 (2017) 522--537.
\newblock \href {http://arxiv.org/abs/1706.08848} {\path{arXiv:1706.08848}},
  \href {http://dx.doi.org/10.1016/j.nuclphysb.2017.05.011}
  {\path{doi:10.1016/j.nuclphysb.2017.05.011}}.

\bibitem{Potapov:2016pgr}
A.~A. Potapov, R.~N. Izmailov, K.~K. Nandi, {Mass decomposition of SLACS lens
  galaxies in Weyl conformal gravity}, Phys. Rev. D93~(12) (2016) 124070.
\newblock \href {http://arxiv.org/abs/1607.01961} {\path{arXiv:1607.01961}},
  \href {http://dx.doi.org/10.1103/PhysRevD.93.124070}
  {\path{doi:10.1103/PhysRevD.93.124070}}.

\bibitem{Cattani:2013dla}
C.~Cattani, M.~Scalia, E.~Laserra, I.~Bochicchio, K.~K. Nandi, {Correct light
  deflection in Weyl conformal gravity}, Phys. Rev. D87~(4) (2013) 047503.
\newblock \href {http://arxiv.org/abs/1303.7438} {\path{arXiv:1303.7438}},
  \href {http://dx.doi.org/10.1103/PhysRevD.87.047503}
  {\path{doi:10.1103/PhysRevD.87.047503}}.

\bibitem{Jizba:2014taa}
P.~Jizba, H.~Kleinert, F.~Scardigli, {Inflationary cosmology from quantum
  Conformal Gravity}, Eur. Phys. J. C75~(6) (2015) 245.
\newblock \href {http://arxiv.org/abs/1410.8062} {\path{arXiv:1410.8062}},
  \href {http://dx.doi.org/10.1140/epjc/s10052-015-3441-6}
  {\path{doi:10.1140/epjc/s10052-015-3441-6}}.

\bibitem{Ghodsi:2014hua}
A.~Ghodsi, B.~Khavari, A.~Naseh, {Holographic Two-Point Functions in Conformal
  Gravity}, JHEP 01 (2015) 137.
\newblock \href {http://arxiv.org/abs/1411.3158} {\path{arXiv:1411.3158}},
  \href {http://dx.doi.org/10.1007/JHEP01(2015)137}
  {\path{doi:10.1007/JHEP01(2015)137}}.

\bibitem{Maldacena:2011mk}
J.~Maldacena, {Einstein Gravity from Conformal Gravity}\href
  {http://arxiv.org/abs/1105.5632} {\path{arXiv:1105.5632}}.

\bibitem{Anastasiou:2016jix}
G.~Anastasiou, R.~Olea, {From conformal to Einstein Gravity}, Phys. Rev.
  D94~(8) (2016) 086008.
\newblock \href {http://arxiv.org/abs/1608.07826} {\path{arXiv:1608.07826}},
  \href {http://dx.doi.org/10.1103/PhysRevD.94.086008}
  {\path{doi:10.1103/PhysRevD.94.086008}}.

\bibitem{tHooft:2016uxd}
G.~'t~Hooft, {Local conformal symmetry in black holes, standard model, and
  quantum gravity}, Int. J. Mod. Phys. D26~(03) (2016) 1730006.
\newblock \href {http://dx.doi.org/10.1142/S0218271817300063}
  {\path{doi:10.1142/S0218271817300063}}.

\bibitem{Flanagan:2006ra}
E.~E. Flanagan, {Fourth order Weyl gravity}, Phys. Rev. D74 (2006) 023002.
\newblock \href {http://arxiv.org/abs/astro-ph/0605504}
  {\path{arXiv:astro-ph/0605504}}, \href
  {http://dx.doi.org/10.1103/PhysRevD.74.023002}
  {\path{doi:10.1103/PhysRevD.74.023002}}.

\bibitem{Diaferio:2008gh}
A.~Diaferio, L.~Ostorero, {X-ray clusters of galaxies in conformal gravity},
  Mon. Not. Roy. Astron. Soc. 393 (2009) 215.
\newblock \href {http://arxiv.org/abs/0808.3707} {\path{arXiv:0808.3707}},
  \href {http://dx.doi.org/10.1111/j.1365-2966.2008.14205.x}
  {\path{doi:10.1111/j.1365-2966.2008.14205.x}}.

\bibitem{Pireaux:2004id}
S.~Pireaux, {Light deflection in Weyl gravity: Critical distances for photon
  paths}, Class. Quant. Grav. 21 (2004) 1897--1913.
\newblock \href {http://arxiv.org/abs/gr-qc/0403071}
  {\path{arXiv:gr-qc/0403071}}, \href
  {http://dx.doi.org/10.1088/0264-9381/21/7/011}
  {\path{doi:10.1088/0264-9381/21/7/011}}.

\bibitem{Elizondo:1994vh}
D.~Elizondo, G.~Yepes, {Can conformal Weyl gravity be considered a viable
  cosmological theory?}, Astrophys. J. 428 (1994) 17--20.
\newblock \href {http://arxiv.org/abs/astro-ph/9312064}
  {\path{arXiv:astro-ph/9312064}}, \href {http://dx.doi.org/10.1086/174214}
  {\path{doi:10.1086/174214}}.

\bibitem{Mannheim:2009qi}
P.~D. Mannheim, {Comprehensive Solution to the Cosmological Constant,
  Zero-Point Energy, and Quantum Gravity Problems}, Gen. Rel. Grav. 43 (2011)
  703--750.
\newblock \href {http://arxiv.org/abs/0909.0212} {\path{arXiv:0909.0212}},
  \href {http://dx.doi.org/10.1007/s10714-010-1088-z}
  {\path{doi:10.1007/s10714-010-1088-z}}.

\bibitem{Mannheim1990}
P.~D. {Mannheim}, {Conformal cosmology with no cosmological constant}, General
  Relativity and Gravitation 22 (1990) 289--298.
\newblock \href {http://dx.doi.org/10.1007/BF00756278}
  {\path{doi:10.1007/BF00756278}}.

\bibitem{Caprini:2018oqe}
C.~Caprini, P.~H{\"{o}}lscher, D.~J. Schwarz, {Astrophysical Gravitational
  Waves in Conformal Gravity}\href {http://arxiv.org/abs/1804.01876}
  {\path{arXiv:1804.01876}}.

\bibitem{Nieuwenhuizen1982}
S.~C. Lee, P.~van Nieuwenhuizen,
  \href{https://link.aps.org/doi/10.1103/PhysRevD.26.934}{Counting of states in
  higher-derivative field theories}, Phys. Rev. D 26 (1982) 934--937.
\newblock \href {http://dx.doi.org/10.1103/PhysRevD.26.934}
  {\path{doi:10.1103/PhysRevD.26.934}}.
\newline\urlprefix\url{https://link.aps.org/doi/10.1103/PhysRevD.26.934}

\bibitem{Isaacson:1967zz}
R.~A. Isaacson, {Gravitational Radiation in the Limit of High Frequency. I. The
  Linear Approximation and Geometrical Optics}, Phys. Rev. 166 (1967)
  1263--1271.
\newblock \href {http://dx.doi.org/10.1103/PhysRev.166.1263}
  {\path{doi:10.1103/PhysRev.166.1263}}.

\bibitem{Isaacson:1968zza}
R.~A. Isaacson, {Gravitational Radiation in the Limit of High Frequency. II.
  Nonlinear Terms and the Ef fective Stress Tensor}, Phys. Rev. 166 (1968)
  1272--1279.
\newblock \href {http://dx.doi.org/10.1103/PhysRev.166.1272}
  {\path{doi:10.1103/PhysRev.166.1272}}.

\bibitem{Saffer:2017ywl}
A.~Saffer, N.~Yunes, K.~Yagi, {The gravitational wave stress¨Cenergy
  (pseudo)-tensor in modified gravity}, Class. Quant. Grav. 35~(5) (2018)
  055011.
\newblock \href {http://arxiv.org/abs/1710.08863} {\path{arXiv:1710.08863}},
  \href {http://dx.doi.org/10.1088/1361-6382/aaa7de}
  {\path{doi:10.1088/1361-6382/aaa7de}}.

\bibitem{CambridgeGeneral}
G.~P.~E. M.~P.~Hobson, A.~N. Lasenby, General relativity an introduction for
  physicists.

\end{thebibliography}

\end{document}